\def\draftversion{false}

\RequirePackage{ifthen}
\ifthenelse{\equal{\draftversion}{true}}{
  \documentclass[aps,prb,10pt,galley,amsmath,amssymb,showpacs,letterpaper]{revtex4-1}
}{
  \documentclass[aps,prb,10pt,twocolumn,amsmath,amssymb,showpacs,letterpaper]{revtex4-1}
}

\usepackage{graphicx}
\usepackage{epsfig}
\usepackage{dcolumn}
\usepackage{bm}
\usepackage{hyperref}

\hypersetup{backref=true, pdfnewwindow=true, colorlinks=true,linkcolor=blue, anchorcolor=blue, citecolor=blue, filecolor=blue, menucolor=blue, urlcolor=blue}

\usepackage[usenames,dvipsnames]{color}

\ifthenelse{\equal{\draftversion}{true}}{
  \marginparwidth 2.7in
  \marginparsep 0.5in
  \newcounter{comm} 
  \def\commnext{\stepcounter{comm}}
  \def\commtext{{\bf\color{blue}[\arabic{comm}]}}
  \def\commmar{{\bf\color{blue}[\arabic{comm}]}}
  \def\dvm#1{\commnext\marginpar{\small DV\commmar: #1}\commtext}
  \def\mtm#1{\commnext\marginpar{\small MT\commmar: #1}\commtext}
  \def\jhm#1{\commnext\marginpar{\small JH\commmar: #1}\commtext}
  \def\vsm#1{\commnext\marginpar{\small VS\commmar: #1}\commtext}
  \def\mlab#1{\marginpar{\small\bf #1}}
  
}{
  \def\dvm#1{}
  \def\mtm#1{}
  \def\jhm#1{}
  \def\vsm#1{}
  \def\mlab#1{}
  
}

\def\beq{\begin{equation}}
\def\eeq{\end{equation}}

\begin{document}

\title{Bloch-type Domain Walls in Rhombohedral BaTiO$_3$}

\author{Maryam Taherinejad}
\email{mtaheri@physics.rutgers.edu}

\affiliation{Department of Physics and Astronomy, Rutgers University,
Piscataway, New Jersey 08854-0849, USA}

\author{David Vanderbilt}

\affiliation{Department of Physics and Astronomy, Rutgers University,
Piscataway, New Jersey 08854-0849, USA}

\author{Pavel Marton}

\affiliation{Institute of Physics, Academy of Sciences of the
Czech Republic, Na Slovance, Czech Republic}

\author{Vilma Stepkova}

\affiliation{Institute of Physics, Academy of Sciences of the
Czech Republic, Na Slovance, Czech Republic}

\author{Jiri Hlinka}

\affiliation{Institute of Physics, Academy of Sciences of the
Czech Republic, Na Slovance, Czech Republic}
\date{\today}
\begin{abstract}
Ferroelectric domain walls (FDWs) are usually considered to
be of Ising type, but there have been suggestions in recent
years that Bloch-type FDWs are also possible in some cases,
e.g., in the rhombohedral phase of BaTiO$_3$.  The mechanically
compatible and electrically neutral FDWs in rhombohedral BaTiO$_3$
are of 71$^\circ$, 109$^\circ$, and 180$^\circ$ type.  We have
investigated these FDWs based both on first-principles calculations
and on a Ginzburg-Landau-Devonshire (GLD) model [P. Marton,
I. Rychetsky, and J. Hlinka, Phys. Rev. B \textbf{81}, 144125
(2010)]. The results from both approaches confirm the Ising nature
of the 71$^\circ$ FDW and the Bloch nature of the 180$^\circ$ FDW,
and predict both Ising-type and Bloch-type FDWs are possible
for the 109$^\circ$ case. 
Considering the relatively small rhombohedral strain in BaTiO$_3$,
the competition between the energies of Bloch and Ising FDWs
can be discussed in terms of a picture in which a Bloch wall is
regarded as being composed of a pair of smaller-angle Ising ones.
A reduction by 40\% in the parameters describing the gradient
term in the GLD model brings it into better agreement with the
first-principles results for detailed properties such as the
energies and widths of the FDWs.

\end{abstract}


\maketitle

\section{Introduction}
\label{sec:intro}

Ferroelectrics find many industrial and commercial
applications, such as in high-dielectric constant capacitors,
ferroelectric thin-film memories, piezoelectric transducers, nonlinear
optical devices, and switches.
The performance of many kinds of ferroelectric devices is affected by
the ferroelectric domain structure and the properties of the domain
boundaries. For example, a recent experiment has shown that the
observed dielectric permittivity of a BaTiO$_3$ single
crystal in the rhombohedral phase varies depending on the
domain structures induced by pre-treatments at higher
temperatures.\cite{wang} Such influences on the mechanical
and electrical properties of devices has motivated theoretical
and experimental work directed toward obtaining a better understanding
of ferroelectric domain structures.

Ferroelectric domain walls (FDWs) are usually considered to be of
Ising type, in which ${\bf P}_\parallel$, the projection of the
polarization vector onto the plane of the domain wall, simply
reverses itself by passing through zero along a high-symmetry
path as one scans through the domain wall. Ising FDWs tend
to be favored because ferroelectrics are generally strongly
electrostrictive, so that a rotation of ${\bf P}_\parallel$ away
from this high-symmetry path would entail a significant elastic
energy cost.
In contrast, the spontaneous magnetostriction which couples the
magnetization and lattice strain in ferromagnetic materials is
typically much weaker.  As a result, magnetic domain walls are
usually much wider, on the order of microns, and the
magnetization vector can rotate away from the high-symmetry path.
The domain wall is denoted as a Bloch or N\'{e}el wall depending
on whether this rotation occurs in a plane parallel or normal
to the domain wall, respectively.

In recent years, however,
there have been some theoretical predictions of the presence of Bloch
and even N\'{e}el components in some ferroelectric
materials and heterostructures.\cite{Meyer, Lee, Behra} In particular,
it has been predicted, in the framework of a phenomenological
Ginzburg-Landau-Devonshire (GLD) model, that the 180$^{\circ}$
FDWs in rhombohedral BaTiO$_3$ should be of Bloch type.\cite{Marton}
This work has motivated us to test whether this behavior is also
reproduced by first-principles density-functional calculations
on BaTiO$_3$.

\begin{figure}
\includegraphics[width=1.8in]{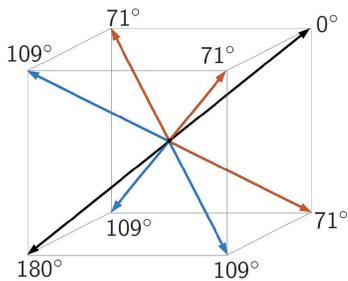}
\caption{\label{fig:p-dir}
(Color online)
The directions of the symmetry-allowed spontaneous polarizations
in rhombohedral BaTiO$_3$. Angles are relative to the reference
direction labeled as 0$^\circ$.}
\end{figure}

Since the discovery of ferroelectricity in this material in
1954,\cite{1954} BaTiO$_3$ has been very widely studied and
has emerged as a kind of prototypical ferroelectric compound.  It
undergoes a sequence of phase transitions from a high-temperature
paraelectric cubic phase to ferroelectric tetragonal, orthorhombic,
and finally rhombohedral phases as the temperature is reduced.
Here we are interested in the zero-temperature
rhombohedral phase, in which the spontaneous polarization prefers
to lie in eight energetically equivalent directions, as
shown by the arrows in Fig.~\ref{fig:p-dir}.  The figure also shows
the possible rotation angles between the spontaneous polarization
directions on the two sides of the domain wall (relative
to the arrow marked as `0'). In the low-temperature rhombohedral
phase of BaTiO$_3$,
the FDWs are therefore of three types: R71$^\circ$,
R109$^\circ$, and R180$^\circ$.
(The `R' denotes a FDW in the rhombohedral phase, following
the notation of Ref.~\onlinecite{Marton}.)
Taking into account the constraints of electrical neutrality
and mechanical compatibility, the plane of the FDW is normal to the
sum of the two polarization vectors for the R71$^\circ$ and
R109$^\circ$ cases, while for the 180$^\circ$ case it can be
either \{$\bar{2}$11\} or \{1$\bar{1}$0\}.

In this work the R71$^\circ$, R109$^\circ$, and R180$^\circ$\{1$\bar{1}$0\}
FDWs in BaTiO$_3$ are investigated using first-principles
calculations in the context of density-functional theory.
While first-principles calculations have been successfully
applied to study FDWs in many other
cases,\cite{Behra,Meyer, Padilla, Lubk}
we are not aware of any previous application to the case of the
rhombohedral phase of BaTiO$_3$.
Our calculations confirm the
Bloch nature of the R180$^\circ$\{1$\bar{1}$0\} FDW and the
Ising nature of the R71$^\circ$ FDW, in agreement with the predictions
of the GLD model. The energy difference between the Bloch R109$^\circ$ and
Ising R109$^\circ$ FDW is very small, which suggests that both types
of the domain wall are possible.
The Bloch walls can be regarded as being
composed of two Ising walls of smaller angles.  A comparison of the sum of
the energies of these constituting Ising walls with that of an Ising-type
solution can explain why the polarization vector picks a Bloch-type path
in some FDWs.  A quantitative comparison of our first-principles
results with those of the GLD model suggests that a 40\% reduction in the
size of a gradient term in the GLD model is needed to bring the
two theories into good agreement.

The manuscript is organized as follows.  In Sec.~\ref{sec:approach}
we describe the geometry of each of the FDWs to be studied. We
also review the first-principles and GLD model approaches
which are used to study the FDWs, and give the details of the
methods used for the first-principles calculations.  The results
from the first-principles calculations and their comparison to
the GLD model are then described in Sec.~\ref{sec:result}. In
Sec.~\ref{sec:discuss} we discuss the competition between Ising
and Bloch configurations in terms of energy considerations,
and in Sec.~\ref{sec:summary} we briefly summarize and discuss
future prospects.

\section{Domain wall and supercell geometries}
\label{sec:geom}

\begin{figure}
\includegraphics[width=2.5in]{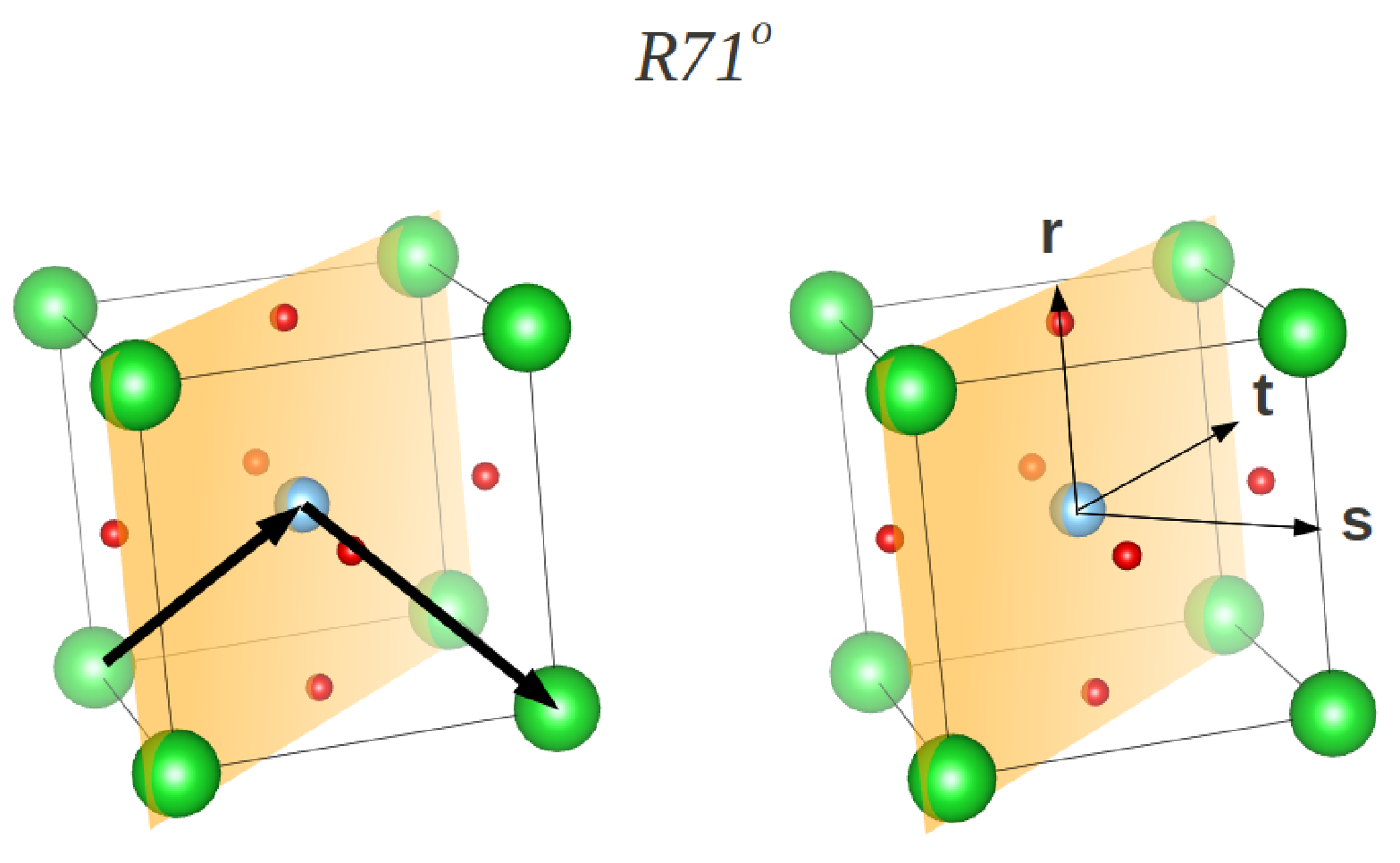}
\includegraphics[width=2.5in]{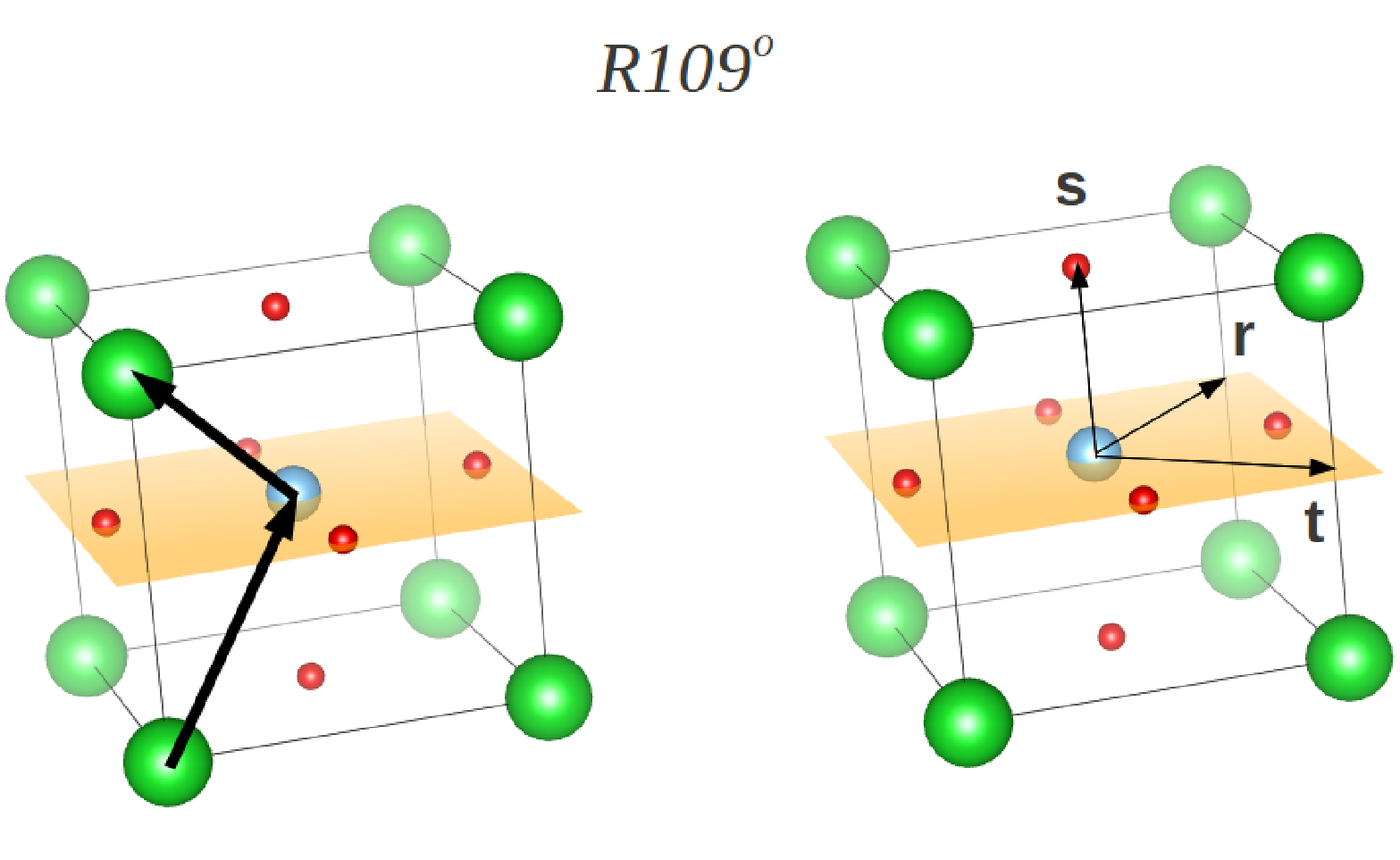}
\includegraphics[width=2.5in]{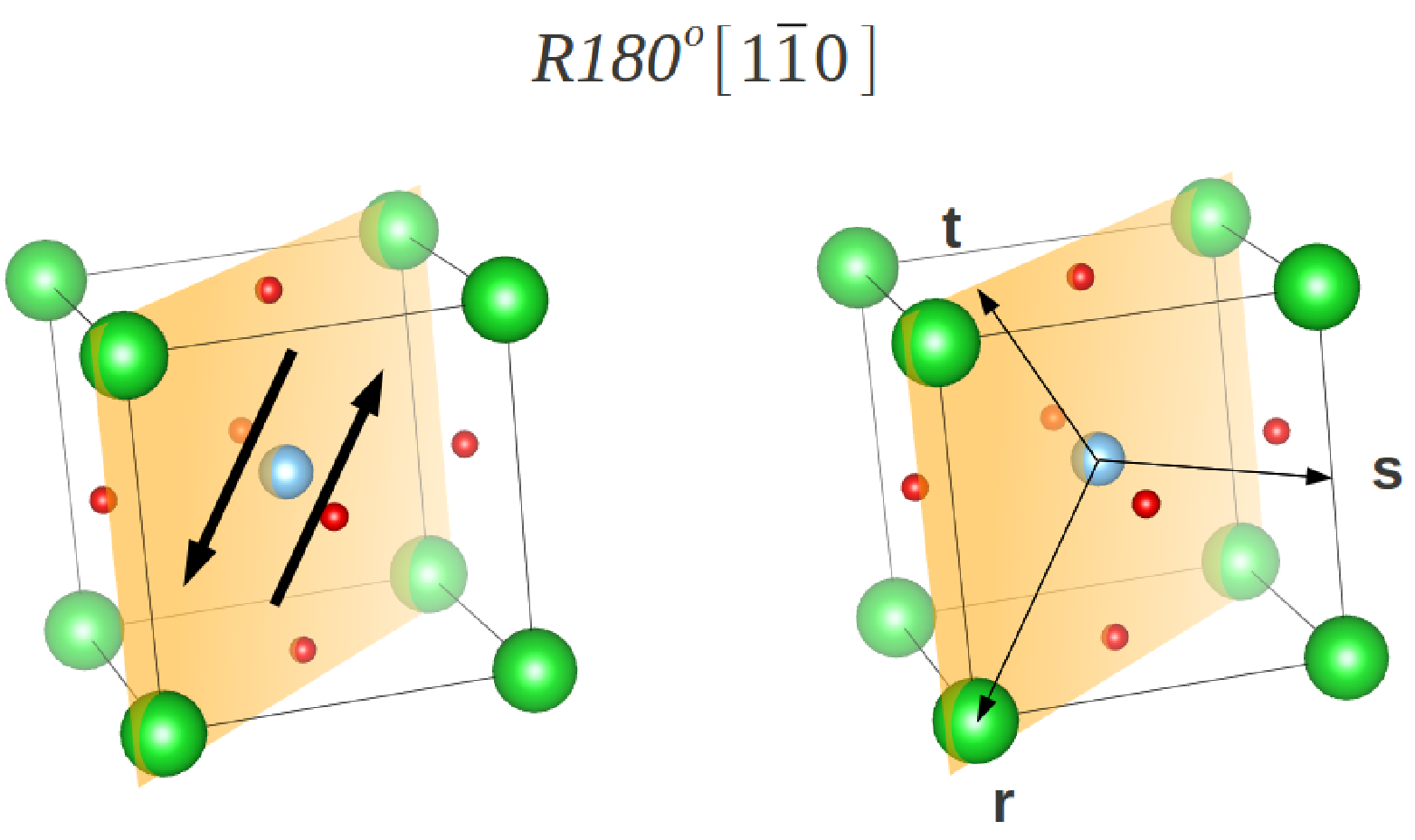}
\caption{\label{fig:unit}
(Color online)
The  R71$^\circ$, R109$^\circ$ and R180$^\circ$\{1$\bar{1}$0\}
FDWs in BaTiO$_3$. The arrows in the figures at left show the
directions of the polarization vectors on the two sides of the FDW,
while those on the right depict the associated symmetry-adapted
coordinate system $(r,s,t)$.}
\end{figure}

The mechanically compatible and electrically neutral FDWs investigated
in this paper are shown in the left column of Fig.~\ref{fig:unit},
where the arrows indicate the orientation of the polarization vectors
$\textbf{P}(-\infty)$ and $\textbf{P}(\infty)$
on the two sides of the domain wall.
In the right column, the symmetry-adapted coordinate system $(r,s,t)$
is shown for each of these walls. The unit vector normal to the wall is
denoted by $\textbf{s}$. The second unit vector $\textbf{r}$ is
chosen to be parallel to $\textbf{P}(\infty)-\textbf{P}(-\infty)$,
the difference between the spontaneous polarizations on the two sides of
the wall; electrically neutrality implies that this is
normal to \textbf{s}.  The third basis vector
is defined as $\textbf{t} = \textbf{r}\times\textbf{s}$.

For the application of the GLD continuum approach, atomistic
details are not important, and specific atomistic geometries do
not have to be considered.  This is obviously not the case, however,
for the first-principles calculations.
These are set up by considering a supercell that is extended
along the direction \textbf{s} normal to the wall, keeping
minimal dimensions in the orthogonal directions.
Ideally we would prefer a supercell containing only a single domain
wall, but this is incompatible with periodic boundary conditions.
Thus, we use supercells containing two equivalent FDWs, where we
enforce this equivalence by imposing
a two-fold screw symmetry (two-fold rotation about
\textbf{s} followed by a half superlattice-vector translation
along the stacking direction \textbf{s}).  If these walls
are well separated, their effect on each other should be small
and the physical properties of the walls should be unaffected.

\begin{figure}
\includegraphics[width=1.6in]{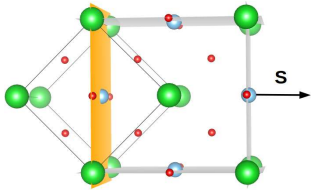}
\caption{\label{fig:unit180}
(Color online)
10-atom rotated building block that is stacked along \textbf{s}
to construct supercells for studying R71$^\circ$ and
R180$^\circ$\{1$\bar{1}$0\} FDWs, which lie normal to \textbf{s}.}
\end{figure}

\begin{figure}
\includegraphics[width=3.4in]{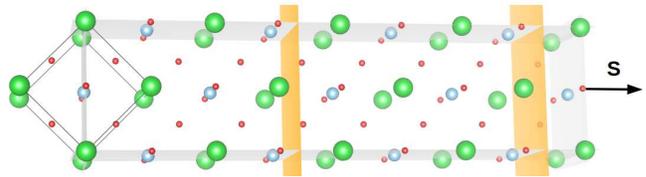}
\caption{\label{fig:suprcel}
(Color online)
A supercell with four rotated 10-atom units stacked in the
\textbf{s} direction. The centers of the FDWs are shown by
the orange planes.}
\end{figure}

We first construct a reference paraelectric supercell by
identifying a minimal building block having lattice vectors
parallel to \textbf{r}, \textbf{s} and \textbf{t}, and repeating
this block $N$ times along the stacking direction \textbf{s}.
An example of such a building block, used for the
R71$^\circ$ and R180$^\circ$\{1$\bar{1}$0\} cases,
is shown in Fig.~\ref{fig:unit180}, and an
example of a supercell built from it is shown in
Fig.~\ref{fig:suprcel}.
An initial configuration for an Ising FDW is then chosen by shifting
the coordinates of the oxygen atoms along the $\textbf{P}(-\infty)$
direction in the first half of the supercell, and along
$\textbf{P}(\infty)$ in the second half.  For the R71$^\circ$ and
R109$^\circ$ cases this results in a configuration with a mirror
symmetry relating the $-\bf t$ and $\bf t$ directions, and since
this symmetry is preserved by the subsequent relaxation of atomic
coordinates,
the resulting relaxed configuration is guaranteed to be of Ising
type.  To initialize a calculation on a Bloch FDW, we also add
oxygen displacement components along $\bf t$ in one FDW and along
$-\bf t$ in the other (still preserving the screw symmetry), so
that effectively
the displacement vector in the \textbf{r-t} plane
is rotated from the \textbf{r} to the \textbf{t} direction across the wall
before pointing to -\textbf{r} on the other side of the wall.
In both the Ising and Bloch cases, the
\textbf{s} component is left unchanged in the initial configuration,
although of course it may relax later.

We then relax the atomic coordinates until all of the forces
fall below a chosen threshold.  While doing this, we constrain the
two in-plane lattice vectors (i.e., in the plane of the FDW) to remain
fixed, consistent with the relaxed strain state of a single-domain
rhombohedral crystal.  We do this because the physical system we are
trying to model is an isolated FDW between very thick domains, in
which case the bulk elasticity dominates over the interface and
fixes the in-plane strain.  We let the third (long) superlattice
vector relax along with the atomic coordinates during the
minimization.  Finally, the polarization
profile is calculated from the pattern of displacements and
the calculated dynamical effective charges as described
in Sec.~\ref{sec:firstprinciples} below.

For all the investigated FDWs, we can always find an Ising-type
FDW solution if we impose an appropriate symmetry constraint.
To test whether this solution is locally stable, we add small
symmetry-lowering atomic displacements, and check whether
it relaxes back to the Ising solution.  We next try starting
from a Bloch-type configuration, with substantial distortions
similar to those that would be present in well-defined domains
of rhombohedral phase.  Again, if this structure relaxes back
to the Ising one, then we conclude that no Bloch FDW was found,
and the Ising solution is stable.  If we find instead that the
calculation converges to a Bloch-like solution, we compare its
energy with that of the Ising solution (if locally stable)
to determine which is the global solution.

The R71$^\circ$ wall has its normal \textbf{s} in the [1$\bar{1}$0]
direction, so that it lies parallel to the diagonal plane in the
primitive cell as shown in Fig.~\ref{fig:unit}.  Therefore, we
consider the building block of Fig.~\ref{fig:unit180}, which is a
10-atom $\sqrt{2}\times\sqrt{2}\times1$ cell obtained by rotating
by 45$^\circ$ around the $z$ axis with respect to the parent cubic
unit cell. In this rotated cell, the FDW lies in the (100) plane.
The simulation supercell is constructed by stacking these units in
the \textbf{s} direction, as shown in Fig.~\ref{fig:suprcel}. The
initial coordinates for the Ising R71$^\circ$ FDW simulation are then
obtained by shifting the oxygen atoms by about
0.1\,\AA\ along [101] in half the supercell, and along [10$\bar{1}$]
in the other half, in the rotated coordinate system.

The R109$^\circ$ FDW lies in the (100) plane, ${\bf s}=\hat{x}$,
and a supercell can easily be made by stacking the primitive
5-atom rhombohedral cells in this direction. The
initial Ising configuration is then set by displacing the oxygen atoms from the
equilibrium positions along the [111] direction on one side
of the FDW, and along the [1$\bar{1}\bar{1}$] direction on the other side.

The R180$^\circ$\{1$\bar{1}$0\} wall is again parallel to the
diagonal plane in the primitive cell, as for the R71$^\circ$
domain wall. So in this case as well, a supercell is made by
stacking the rotated 10-atom units shown in Fig.~\ref{fig:unit180}
in the \textbf{s} direction. The initial configuration for the
R180$^\circ$\{1$\bar{1}$0\} Ising FDW is obtained by shifting
the oxygen atoms from their equilibrium positions along the [011]
direction on one side of the FDW and along the opposite direction
on the other side, in the rotated coordinate system.

For the Ising R180$^\circ$\{1$\bar{1}$0\} case
there is also the possibility of imposing a higher symmetry by
insisting that the FDW lie exactly in a Ba-Ti-O plane, or exactly
in an O-O plane, which can be accomplished by adopting an
inversion center through one of the atoms lying
in the FDW.  This is not possible for the R71$^\circ$ and
R109$^\circ$ FDWs, since the presence of a global $P_s$ component
makes the ``front'' and ``back'' sides of the FDW distinguishable
and rules out the presence of an inversion symmetry.

\section{Computational Approach}
\label{sec:approach}

\subsection{First-principles calculations}
\label{sec:firstprinciples}

The calculations are done using the ABINIT implementation of
density functional theory\cite{DFT} within the local-density
approximation (LDA) using the Perdew-Zunger exchange-correlation
functional.\cite{LDA}
Ultrasoft pseudo\-po\-ten\-tials,\cite{us-psp0} in which
semicore $s$ and $p$ states are included in the valence for
Ba and Ti, were converted for use as
projector augmented-wave (PAW) potentials\cite{PAW}
using the USPP2ABINIT package.  The plane wave cut-off
and the energy cut-off for the fine FFT grid are set to 25
Ha and 40 Ha respectively. The tolerance on the difference of forces
in successive iterations in a self-consistent-field (SCF) cycle
is set to $5.0\times10^{-10}$ hartree/Bohr, which reached twice
successively, causes one SCF cycle to stop and ions to be moved.
The structural optimizations are done using the
Broyden-Fletcher-Goldfarb-Shanno
minimization.\cite{BFGS}

The supercells employed for studying R71$^\circ$ and
R180$^\circ$ FDWs  are of dimensions
$\sqrt{2}N a \times \sqrt{2}a \times a$, where
$a$ is the primitive lattice constant and $N$
is the number of 10-atom units stacked in the
\textbf{s}-direction. Therefore, a $1 \times 4 \times 6$
Monkhorst-Pack\cite{monkh} $k$-mesh is chosen for
simulating these  domain walls. The supercell for the
R109$^\circ$ FDW is made by stacking the 5-atom primitive
cells in one direction, so a $1 \times 6 \times 6$
Monkhorst-Pack\cite{monkh} $k$-mesh seems a proper
choice for simulating this FDW.

We also compute the dynamical effective charges\cite{zhong}
$Z^*$ in bulk paraelectric cubic BaTiO$_3$.
The dynamical charge tensor $Z^*_{i\alpha,\beta}$ of a given atom
measures the dipole induced along $\beta$ by a displacement of atom
$i$ along $\alpha$.  In many oxides including BaTiO$_3$, these
charges are quite different from the formal ionic charges.
The $Z^*$ tensors are computed by finite differences, i.e., by
making small displacements and calculating the resulting change
in Berry-phase polarization.\cite{Vanderbilt}

These dynamical charges are then used as an ingredient in an
algorithm\cite{Meyer} by which we map out the polarization profiles in
FDW-containing supercells, as follows.
The polarization is only changing along the stacking direction and
is constant in the planes normal to this direction. So first the
contribution from each layer to the polarization in direction $\alpha$
arising from displacements of atoms $j$ in direction $\beta$
is calculated as
\begin{equation}
p^{(l)}_{\alpha}=\sum_{\beta,j\in\l} Z^*_{j\beta,\alpha} \, u_{j\beta}
\end{equation}
where $l$ is a layer index. In the supercell
used for studying the R109$^\circ$ FDW these are
Ba-O and Ti-O-O layers, while for the R71$^\circ$
and R180$^\circ$\{1$\bar{1}$0\} FDWs built from the rotated
10-atom units $l$ refers to Ba-Ti-O and O-O layers.
If we break the supercell into smaller cells centered on these
layers, we can assign a local polarization to cell $l$ by
counting its own contribution and half that of each neighbor,
i.e.,
\begin{equation}
P^{(l)}_{\alpha}=\frac{1}{\Omega} \left( \frac{1}{2}p^{(l-1)}_{\alpha}
   + p^{(l)}_{\alpha} +\frac{1}{2}p^{(l+1)}_{\alpha}\right),
\label{eq:Pl}
\end{equation}
where $\Omega$ is the volume of the cell.

\subsection{The GLD model}
\label{sec:gld}

We review the Ginzburg-Landau-Devonshire model used in
Ref.~\onlinecite{Hlinka}, which is again used here to model
the FDW properties and compare with the first-principles
results.
The excess free energy $F$ relative to the reference cubic paraelectric
state is expressed in terms of polarization and strain fields as
\begin{eqnarray}
\label{eq:fdens}
F[\lbrace P_i,P_{i,j},e_{i,j}\rbrace] = \int f(\textbf{r})\,d\textbf{r} ,
\end{eqnarray}
where $f$ is the GLD free-energy density which is taken to be a
function of the polarization components $P_i$, their spatial
derivatives $P_{i,j}=\partial P_i / \partial x_j$, and
strain components $e_{i,j}$.  In particular, $f$ is expressed
in terms of Landau, elastic, electrostriction, and gradient terms:
\begin{eqnarray}
\label{eq:fdef}
f = f_L^{(e)}\{P_i\}+f_c\{e_{ij}\}+f_q\{P_i,e_{ij}\}+f_G\{P_{i,j}\}.
\end{eqnarray}
In Ref.~\onlinecite{Hlinka}, explicit forms were given for each
of the terms in this expression, parameter values were estimated
from the bulk single-crystal properties of BaTiO$_3$, and the
GLD model was used to investigate domain-wall properties.

Here we are especially concerned with the gradient or Ginzburg terms
in the free-energy expansion, which take the form
\begin{equation}
\begin{split}
f_G = & \frac{1}{2} G_{11}(P_{1,1}^2+P_{2,2}^2+P_{3,3}^2)\\
& +G_{1,2}(P_{1,1}P_{2,2}+P_{2,2}P_{3,3}+P_{1,1}P_{3,3}) \\
& +\frac{1}{2}G_{44}\left[(P_{1,2}+P_{2,1})^2+(P_{2,3}+P_{3,2})^2\right. \\
& \hspace{1.8cm}\left. +(P_{3,1}+P_{1,3})^2\right].
\label{eq:fG}
\end{split}
\end{equation}
As discussed in Ref.~\onlinecite{Hlinka}, considerable caution was
required in extracting the $G$ coefficients from inelastic neutron
scattering experiments, and the remaining uncertainties are
significant. The $G$ tensor has an important effect on
the widths and energies of the FDWs, so that the uncertainties
in the values of these coefficients is a limiting factor in
determining the properties of the FDWs.
The original parameters of Ref.~\onlinecite{Marton} describing the gradient
terms are
$G_{11}=51\times 10^{-11}$\,Jm$^3$C$^{-2}$,
$G_{12}=-2\times 10^{-11}$\,Jm$^3$C$^{-2}$, and
$G_{44}= 2\times 10^{-11}$\,Jm$^3$C$^{-2}$.

In order to establish a better estimate of these coefficients, the
GLD model is employed below to recalculate the polarization profiles
of R71$^\circ$, R109$^\circ$, and R180$^\circ$\{1$\bar{1}$0\} FDWs
at zero temperature using modified $G$ coefficients, and the results
are compared with first-principles ones.  In order to facilitate
the comparisons, these GLD model calculations have been performed
on the identical geometries as in the first-principles calculations.
That is, we impose periodic boundary conditions corresponding to the
size of the first-principles supercells, impose the same two-fold
screw symmetry as was used there, and
specify the strain state to be consistent with infinite domains.

\subsection{Domain-wall width}
\label{sec:width}

In the case of a domain wall whose polarization profile can
be fit to a hyperbolic tangent, $p(x)=p_0\,\tanh(x/\xi)$, a common
definition of the width is $w=2\xi$.  However, some of the domain
walls to be studied here have unusual polarization profiles that
do not resemble a single hyperbolic tangent at all.  To accommodate
such cases, we define $w$ as the width of the region within which
$|p(x)|/p_0<\tanh(1)=0.762$.  This definition has the advantages of
being globally reasonably and of reducing to the conventional
definition above when the FDW does resemble a hyperbolic tangent.
We adopt this definition throughout the remainder of this work.

\section{Results}
\label{sec:result}

\begin{table}
\caption{\label{tab:charge}%
The experimental and theoretical values of the dynamical effective charges of
the Ba, Ti and O atoms in BaTiO$_3$ (in units of the charge quantum $e$). }
\begin{ruledtabular}
\begin{tabular}{lllll}
&\mbox{$Z_{\rm Ba}$}&\mbox{$Z_{\rm Ti}$}&\mbox{$Z_{\rm O}\perp$}&\mbox{$Z_{\rm O}\parallel$}\\
\hline
Nominal ionic&2&4&$-$2&$-$2\\
$Z^*$ (Exp.\cite{Axe})
&2.9&6.7&$-$2.4&$-$4.8\\
$Z^*$ (LDA )
& 2.75 & 7.18 & $-$1.86&$-$5.65 \\
\end{tabular}
\end{ruledtabular}
\end{table}

BaTiO$_{3}$ has been studied
extensively both experimentally and theoretically. Our
first-principles computed values of 3.95\,\AA\ for the lattice
constant and 89.93$^\circ$ for the rhombohedral angle
can be compared with experimental values\cite{Hewat}
of 4.00\,\AA\ and 89.87$^\circ$, respectively.
Our results are consistent with the experience that the LDA
typically gives slightly underestimated values for the unit-cell
volume and ferroelectric distortion.\cite{Rabe}

We have also computed the values of the dynamical charge tensors
for cubic paraelectric BaTiO$_3$, to be used as an ingredient in the
algorithm for computing polarization profiles as described in
Sec.~\ref{sec:firstprinciples}.  The calculated values of $Z^*$ for Ba, Ti and O
in BaTiO$_3$ are compared to the experimental values as well as ionic
charges in Table~\ref{tab:charge}. The values of $Z_{\rm O}\parallel$ and
$Z_{\rm O}\perp$ refer to the $Z^*$ of the oxygen ion when it is displaced
along the Ti-O direction or perpendicular to it, respectively.

A Berry-phase calculation of the polarization\cite{Vanderbilt} using
the relaxed atomic positions in rhombohedral BaTiO$_3$ yields
a value of 30\,$\mu$C/cm$^2$. The polarization calculations using the
atomic displacements and the theoretical $Z^*$ values yields
31\,$\mu$C/cm$^2$, which is slightly underestimated compared
to the experimental value\cite{Hewat} of 33.5\,$\mu$C/cm$^2$
as expected. The GLD model, on the other hand, yields a value of
38$\mu$C/cm$^2$.

In the following subsections, we report the results of our
supercell calculations for the
R71$^\circ$, R109$^\circ$, and R180$^\circ$\{1$\bar{1}$0\}
ferroelectric domain walls.

\subsection{The R71$^\circ$ domain wall}
\label{sec:r71}

\begin{figure}
\includegraphics[width=3.4in]{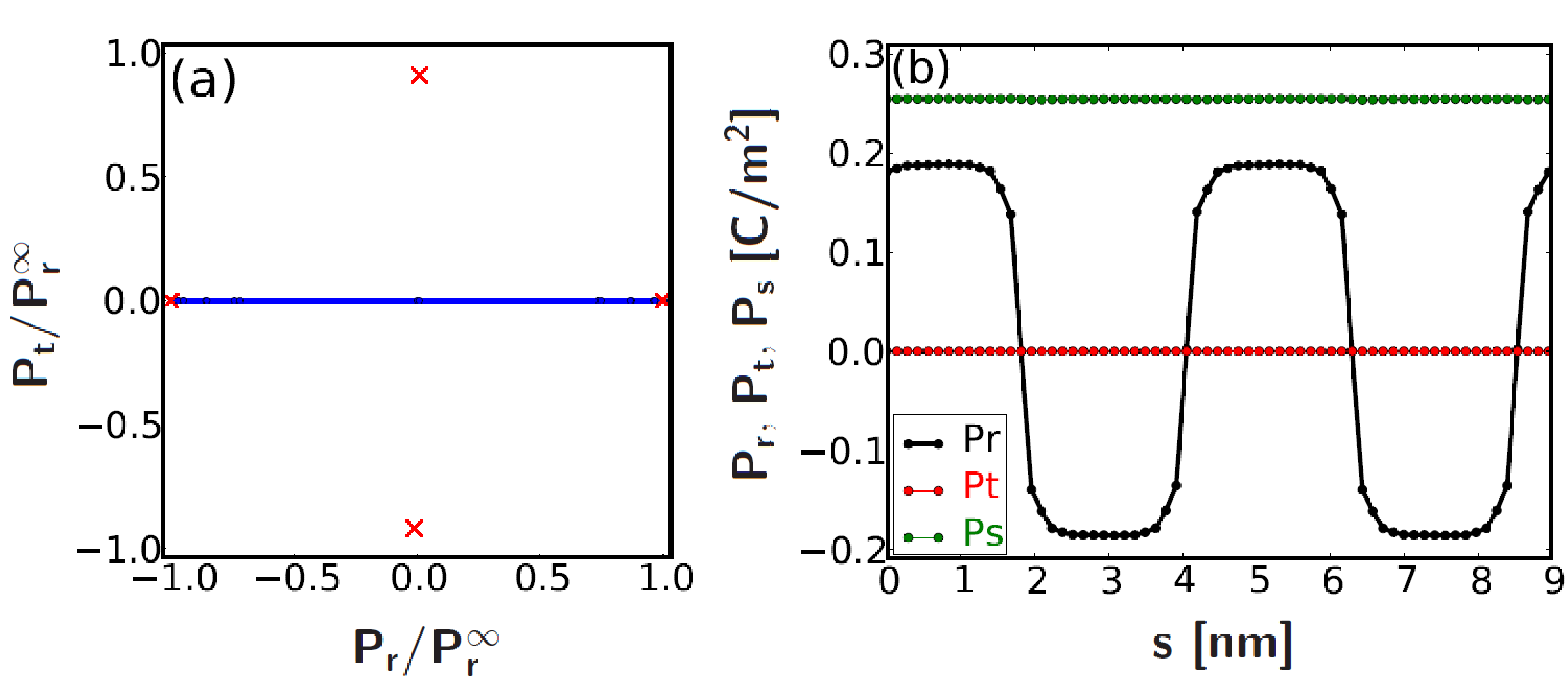}
\includegraphics[width=3.4in]{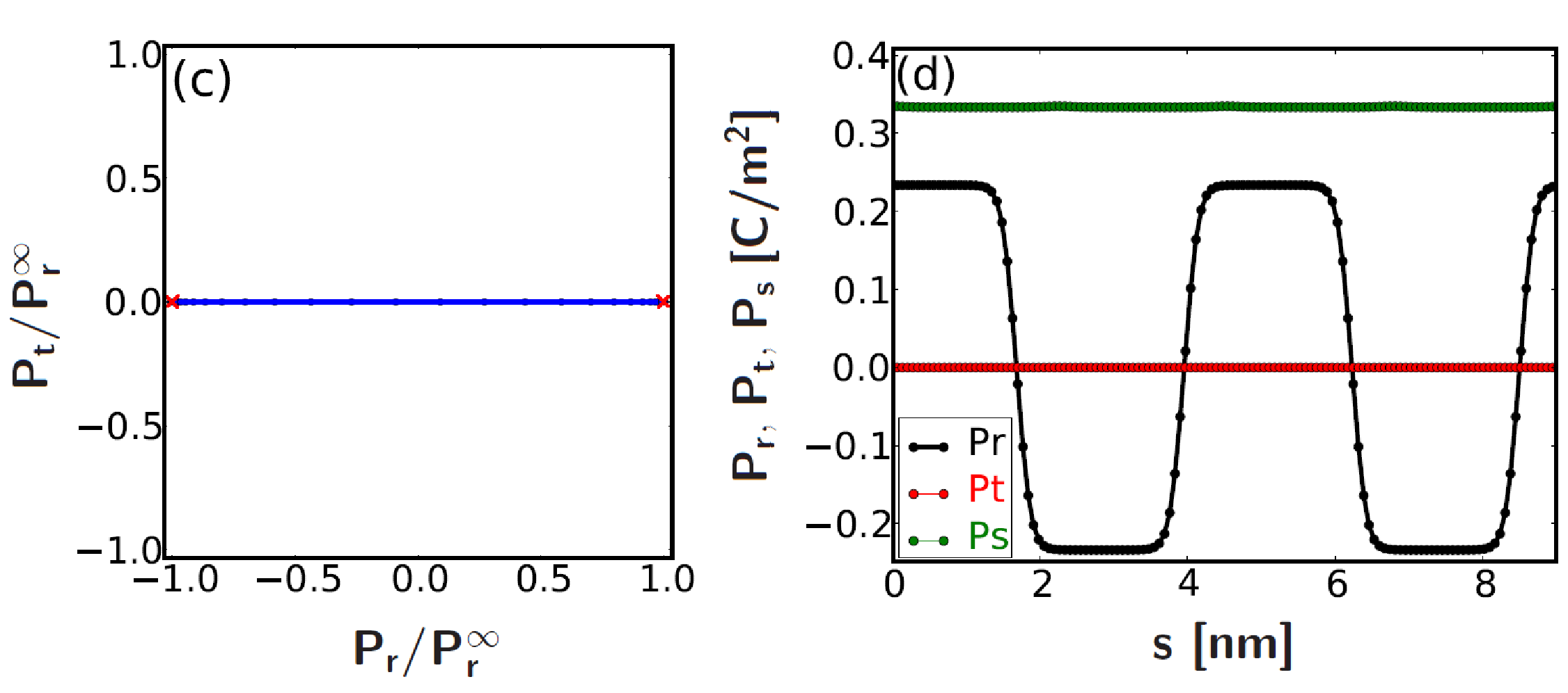}
\caption{\label{fig:71}
(Color online)
Polarization profiles of Ising R71$^\circ$ FDWs in rhombohedral
BaTiO$_3$ as calculated using (a-b) first-principles calculations,
and (c-d) the GLD model.  Panels (a) and (c): parametric plots
showing polarization values in the $P_r$--$P_t$ plane.  Crosses
indicate local energy minima associated with homogeneous
rhombohedral domains.  Panels (b) and (d): polarization
components as a function position $s$ along the supercell direction
(two supercells containing four FDWs are shown for clarity).}
\end{figure}

We have investigated the R71$^\circ$ FDW in rhombohedral BaTiO$_3$
by carrying out first-principle calculations on an 80-atom
supercell made by stacking 10-atom rotated units as described in
Sec.~\ref{sec:geom}.
In this case we only found an Ising FDW; perturbing this by
adding symmetry-lowering components only led back to the Ising structure
upon further relaxation. The polarization profile computed from the
relaxed Ising structure using Eq.~(\ref{eq:Pl}) is displayed in two ways in
Fig.~\ref{fig:71}(a-b). The left panel, Fig.~\ref{fig:71}(a), shows
the $P^{(l)}_r$ and $P^{(l)}_t$ values for each layer $l$, while
Fig.~\ref{fig:71}(b) shows plots of $P^{(l)}_r$, $P^{(l)}_s$ and
$P^{(l)}_t$ as a function of position $s$ while scanning through a
sequence of four domain walls (two entire supercells).
As is clear from this figure,
the $P_t$ component remains zero everywhere in the
supercell, which clearly indicates the Ising nature of this FDW.
The $P_r$ value reverses quite suddenly and attains a value
very close to its saturation bulk value deep inside each
domain, indicating a rather narrow FDW width.
The $P_s$ value is almost exactly constant; as we shall
see, this is true of all FDWs in this study, consistent with the
expectation that inhomogeneities in $P_s$ would result in bound
charge which in turn would involve an extra Coulomb energy cost.

The above results are in good qualitative agreement with those
of the GLD model, but the GLD domain-wall width of 0.58\,nm
is almost twice that of the first-principles prediction of 0.33\,nm.
This suggests the need for a reduction of the zero-temperature
values of the gradient terms in the GLD model to bring it
into better agreement with the first-principles results.
This reduction of the gradient terms will be even more important
in the R180$^\circ$\{1$\bar{1}$0\} case, to be discussed
in Sec.~\ref{sec:r180}.  As explained there, we have
chosen to reduce all three of the gradient coefficients
in Eq.~(\ref{eq:fG}) by 40\% in order to arrive at an improved
GLD model.

The results computed for the R71$^\circ$ FDW using this modified
GLD model are presented in Fig.~\ref{fig:71}(c-d) using the
same plotting conventions as for the first-principles results
in Fig.~\ref{fig:71}(a-b).  The FDW width is now 0.37\,nm.
Furthermore, the GLD domain-wall energy is reduced from 5.0 to
3.2\,mJ/m$^2$, to be compared with the first-principles value
of 3.8\,mJ/m$^2$.
While these numerical values should be interpreted
reservedly in view of the uncertainties in both theories,
it is clear that the GLD theory is in better agreement
with the first-principles theory after the reduction of the
strength of the gradient term.

\subsection{The R109$^\circ$ domain wall}
\label{sec:r109}

\begin{figure}
\includegraphics[width=3.4in]{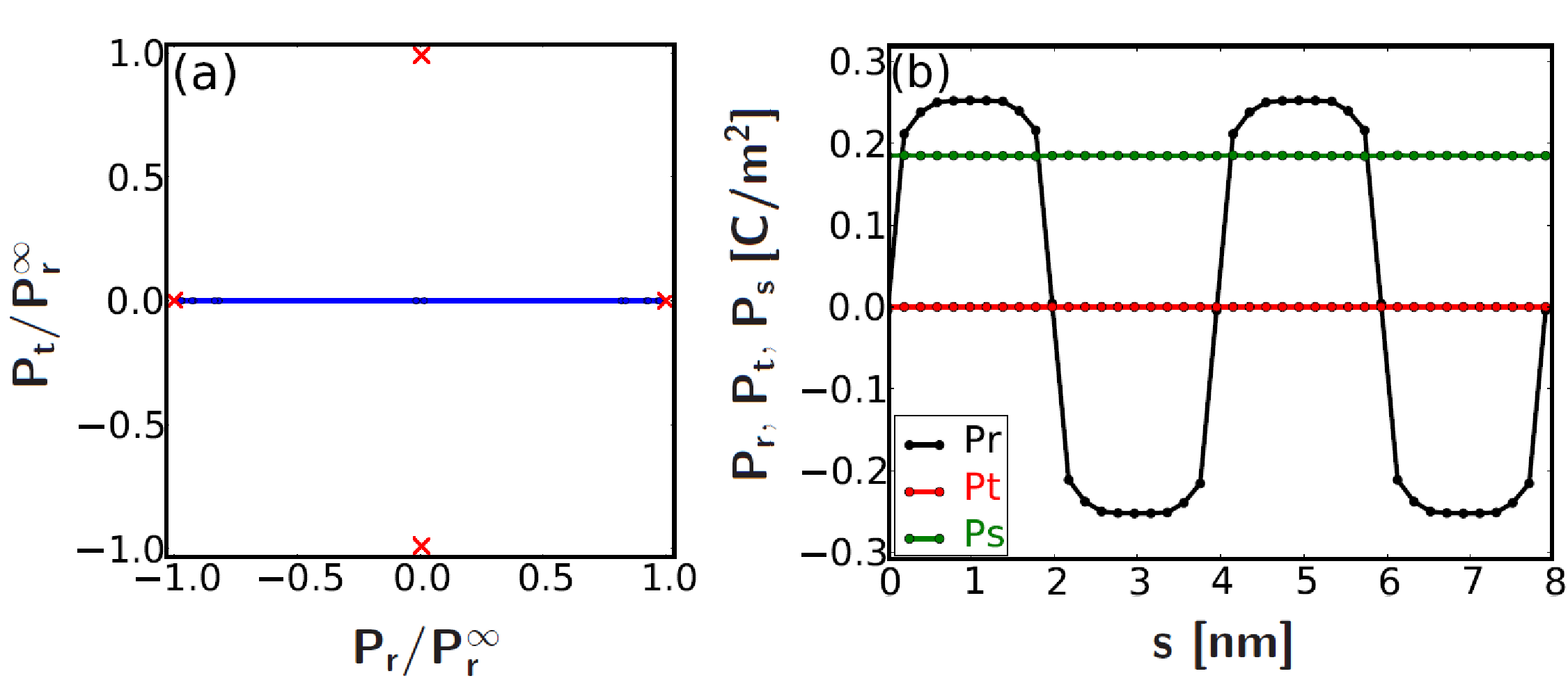}
\includegraphics[width=3.4in]{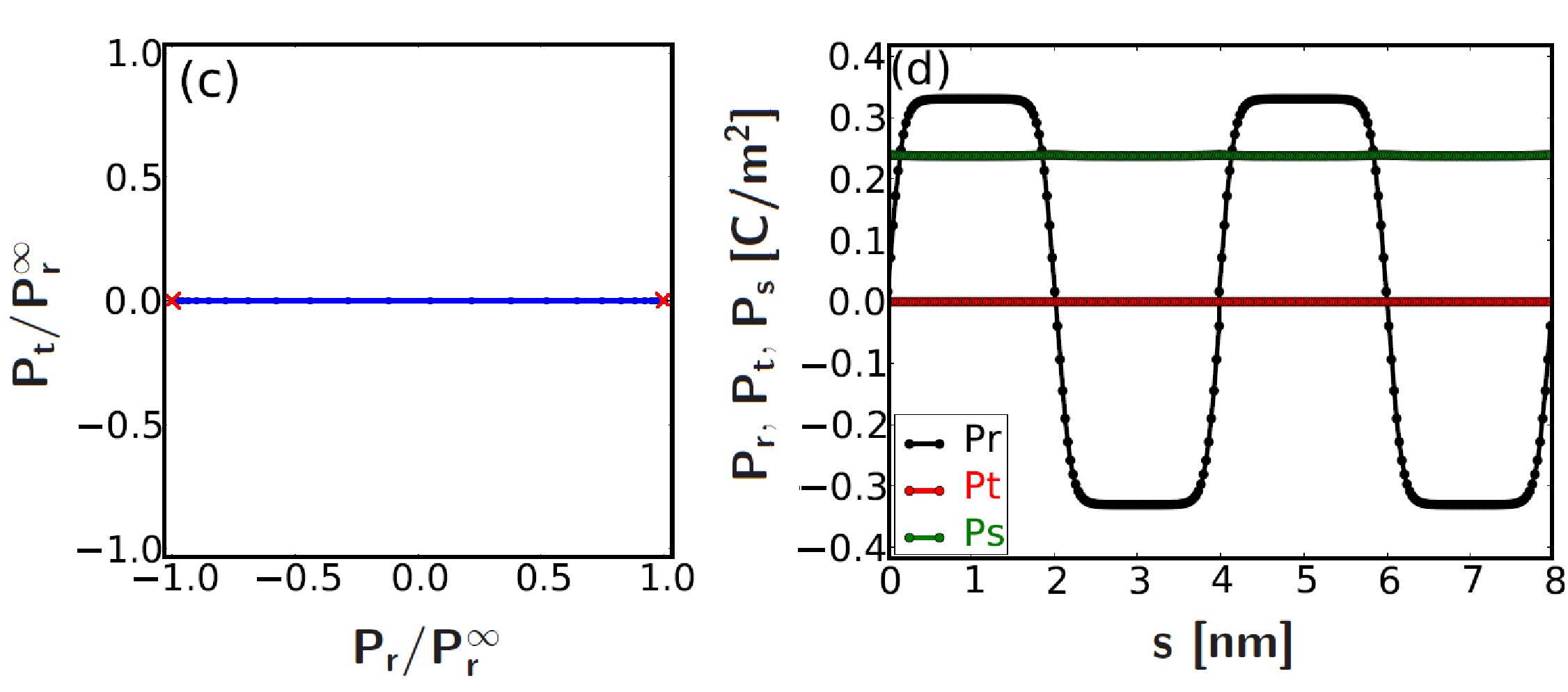}
\caption{\label{fig:I109}
(Color online)
Polarization profiles for Ising R109$^\circ$ FDWs in BaTiO$_3$
using (a-b) first-principles calculations and (c-d) the GLD model.
Details are as in Fig.~\ref{fig:71}.}
\end{figure}

\begin{figure}
\includegraphics[width=3.4in]{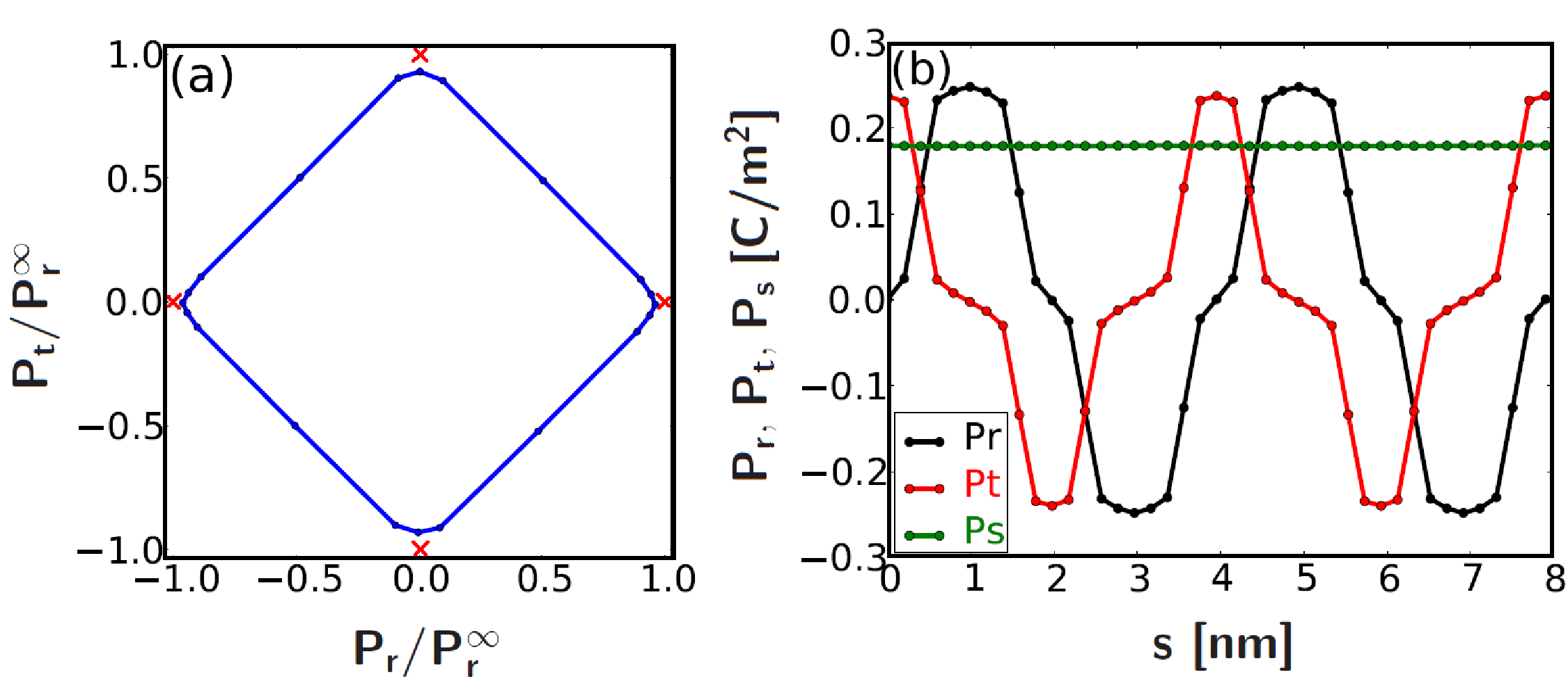}
\includegraphics[width=3.4in]{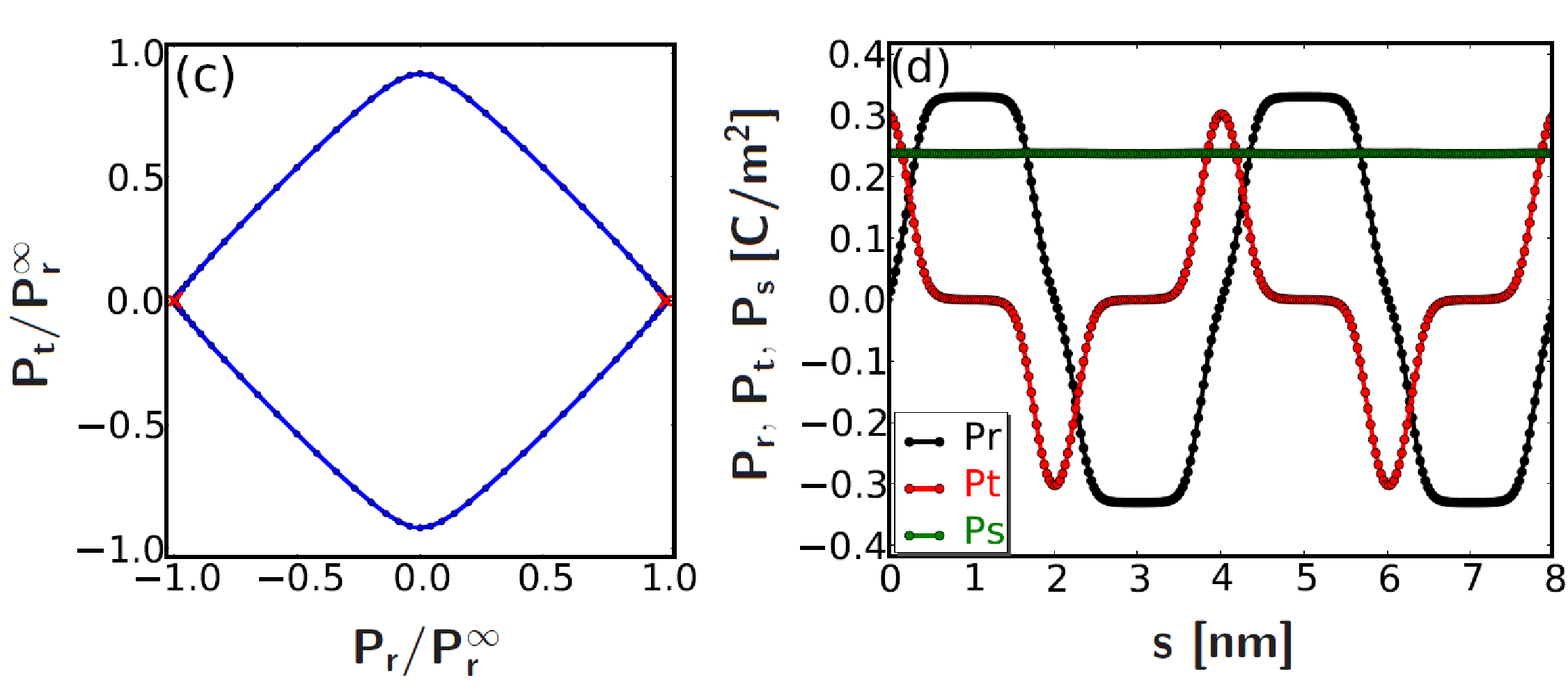}
\caption{\label{fig:B109}
(Color online)
Polarization profiles for Bloch R109$^\circ$ FDWs in BaTiO$_3$
using (a-b) first-principles calculations and (c-d) the GLD model.
Details are as in Fig.~\ref{fig:71}.}
\end{figure}

According to the GLD model calculations of Marton {\it et al.},\cite{Marton}
both Ising and Bloch solutions are possible for the R109$^\circ$ FDW
in BaTiO$_3$. Our first-principles results confirm this picture.

Starting first with the Ising case, Fig.~\ref{fig:I109}(a-b)
shows the polarization profile for a 50-atom supercell in which the
atomic positions have been relaxed from an initial configuration with
two Ising-type R109$^\circ$ FDWs. The domain wall is again fairly
narrow, though not quite as narrow as in the R71$^\circ$ case.
The energy and width of this Ising R109$^\circ$ FDW are calculated
from first principles to be 0.36\,nm and 11.1\,mJ/m$^2$, respectively.
The corresponding GLD results using the reduced gradient term,
shown in Fig.~\ref{fig:I109}(c-d), are clearly in good
qualitative agreement.

When the atomic positions are relaxed from an appropriately
distorted initial configuration, a Bloch-type solution for this
wall is found. The first-principles polarization profiles computed
for the Bloch-type R109$^\circ$ FDW in a 50-atom supercell are
shown in Fig.~\ref{fig:B109}(a-b).  To a first approximation, this
Bloch FDW can be
regarded as a composition of two 71$^\circ$ FDWs in close proximity.
In part for this reason,
the Bloch-type FDW is clearly
broader than the Ising one. However, the energies of the Bloch and
Ising solutions are found to be almost identical, with the Ising
one being only $\sim$2-3\% lower in energy.  If we extrapolate
to larger separations between FDWs we might expect the Bloch
energy to fall more than the Ising one, because of the larger
FDW width in the Bloch case.  This suggests that both types of
R109$^\circ$ FDWs have similar energies and that both might
be found in rhombohedral BaTiO$_3$ crystals.  The reason for
existence of both Ising and Bloch solutions for the R109$^\circ$
FDW is discussed in Sec.~\ref{sec:discuss}.
The modified GLD model again gives good qualitative and semiquantitative
agreement with the first-principles results for the case of the Bloch
R109$^\circ$ FDW, as shown in Fig.~\ref{fig:B109}(c-d).

\subsection{The R180$^\circ$\{1$\bar{1}$0\} domain wall}
\label{sec:r180}

A major result of the GLD study of Marton {\it et al.},\cite{Marton}
was the prediction that the lowest-energy FDW for the
R180$^\circ$\{1$\bar{1}$0\} case in BaTiO$_3$ is of Bloch type,
with an energy lying 10\% lower than that of
its Ising counterpart. Verifying this
result from first principles is computationally more challenging than
studying the R71$^\circ$ or R109$^\circ$ FDWs.
The R180$^\circ$\{1$\bar{1}$0\} FDW has the biggest rotational angle
and is hence the widest of the three investigated FDWs. Moreover,
the Bloch-type FDWs are generally much broader than the Ising-type ones.
Preliminary first-principles calculations showed that supercells smaller
than 80 atoms are too small to accommodate two R180$^\circ$ FDWs; initial
60-atom Bloch-wall supercells relaxed to unreasonable configurations.
We have therefore carried out our calculations on an 80-atom supercell.
While the polarization does not quite have room to reach its
saturation value between neighboring FDWs, at least we obtain a
stable solution that can reveal the Ising or Bloch nature of this FDW.

\begin{figure}
\includegraphics[width=3.4in]{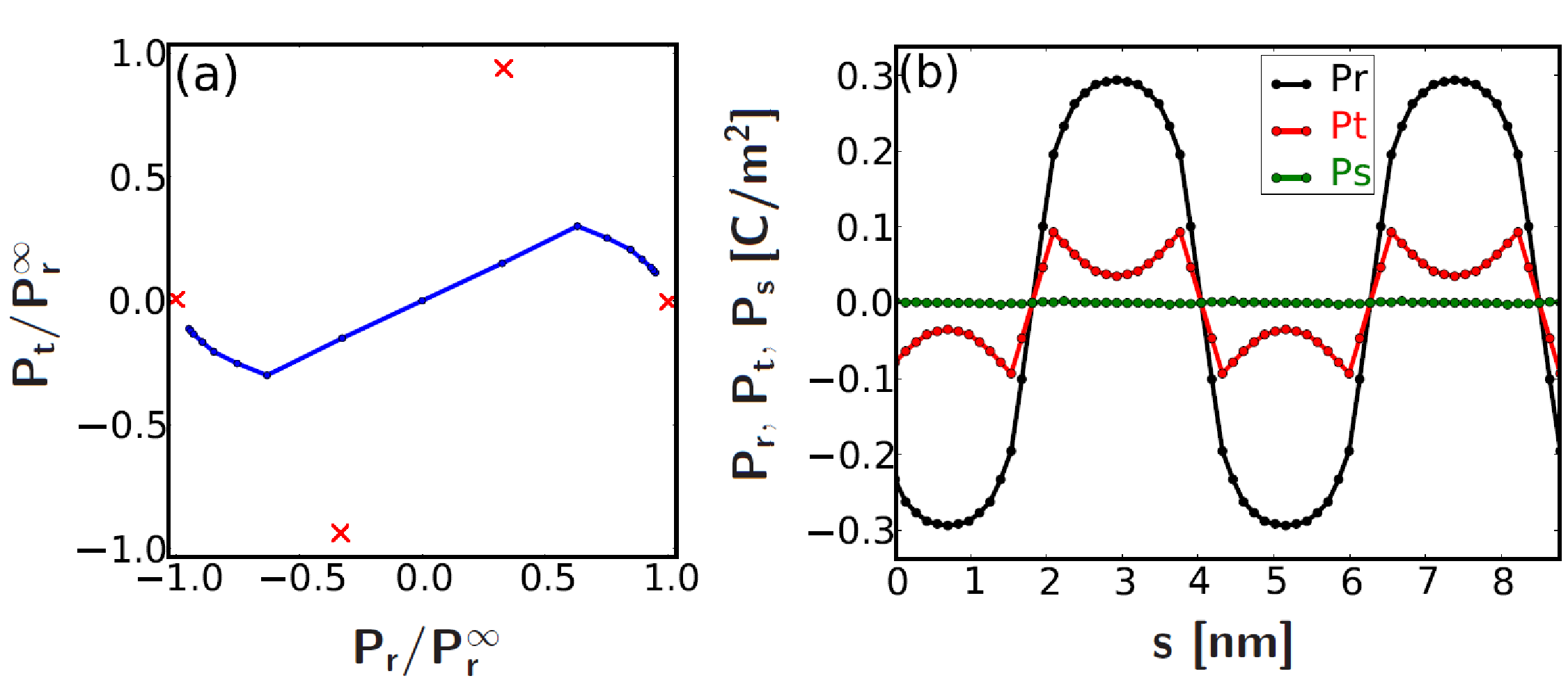}
\includegraphics[width=3.4in]{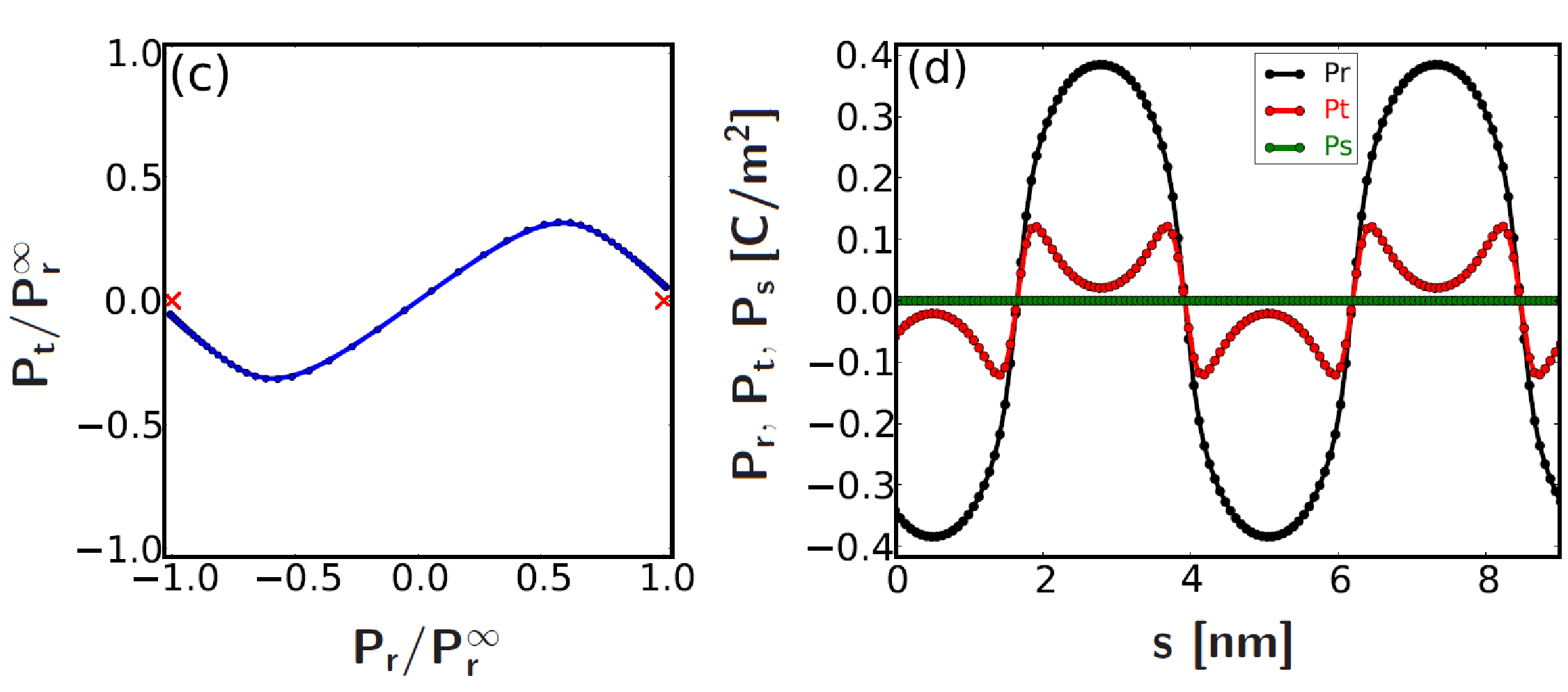}
\caption{\label{fig:180ising}
(Color online)
Polarization profiles for Ising R180$^\circ$\{1$\bar{1}$0\} FDWs
in BaTiO$_3$ using first-principles calculations.
Details are as in Fig.~\ref{fig:71}.}
\end{figure}

As mentioned in the last paragraph of Sec.~\ref{sec:geom},
it is possible to enforce an Ising-like geometry in the
R180$^\circ$\{1$\bar{1}$0\} case by imposing an initial
inversion symmetry about an atom in the center of the FDW and
preserving this symmetry during relaxation.  When we apply
our first-principles calculations including this symmetry
constraint, we arrive at a configuration like that shown in
Fig.~\ref{fig:180ising}(a-b), which is Ising-like in the sense
that $P_r=P_t=0$ in the center of the wall.  However, $P_t$ has
substantial excursions away from zero, with the polarization path
following an S-like curve in $(P_r,P_t)$ space, as can be seen
clearly in Fig.~\ref{fig:180ising}(a).

\begin{figure}
\includegraphics[width=3.4in]{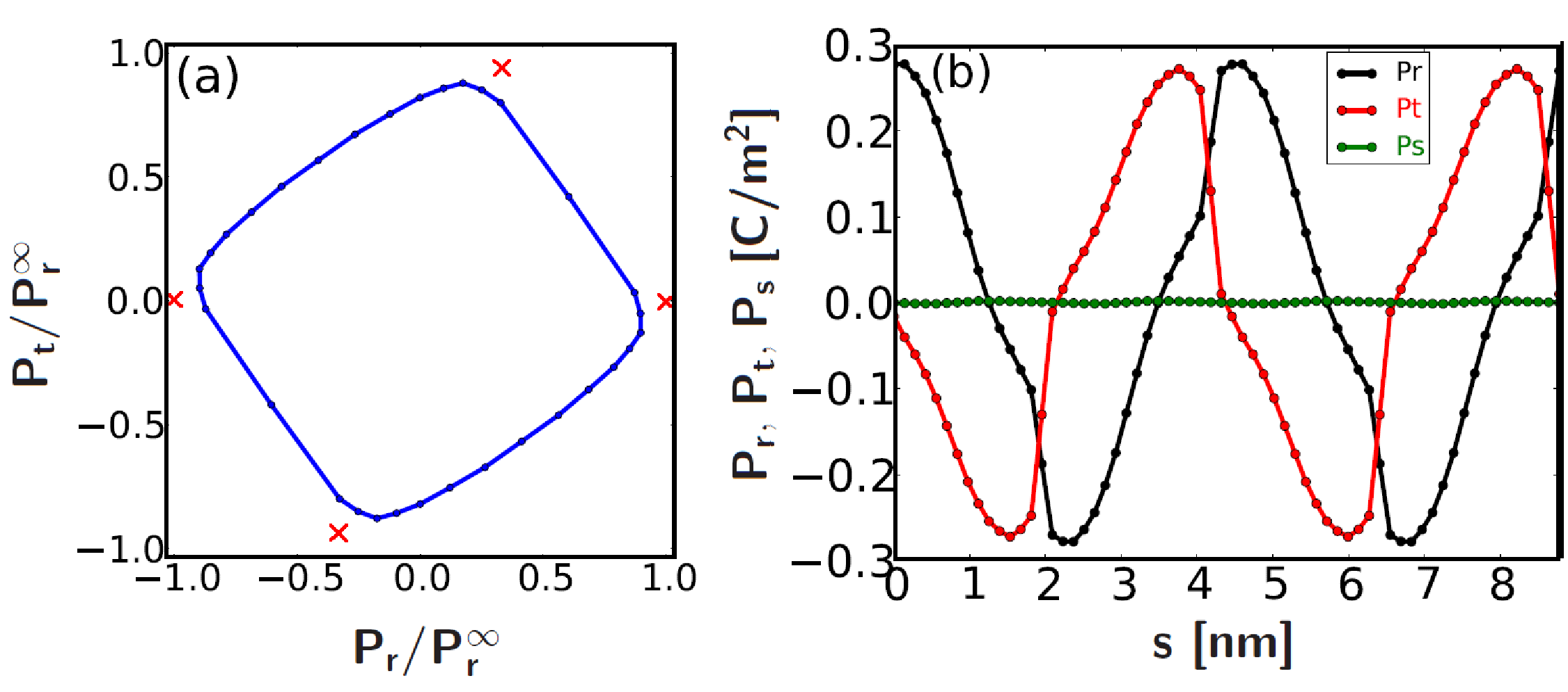}
\includegraphics[width=3.4in]{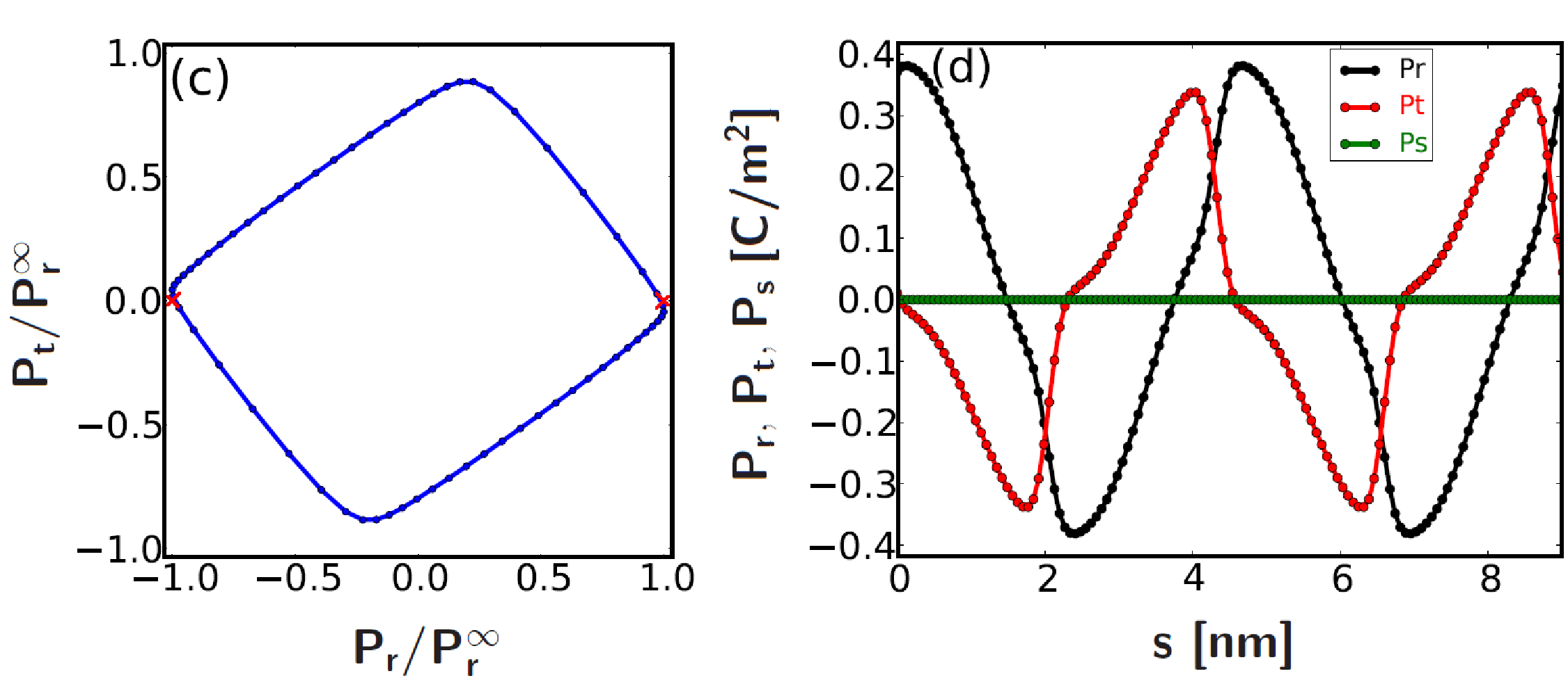}
\caption{\label{fig:180}
(Color online)
Polarization profiles for Bloch R180$^\circ$\{1$\bar{1}$0\}
FDWs in BaTiO$_3$
using (a-b) first-principles calculations and (c-d) the GLD model.
Details are as in Fig.~\ref{fig:71}.}
\end{figure}

However, we find that if we do not impose this special symmetry,
most initial conditions relax to the Bloch configuration shown
in Fig.~\ref{fig:180}(a-b).  Even if we start from an Ising-like
configuration and break the symmetry only slightly, we find
that the simulation will eventually relax to the Bloch configuration.
It is clear from Fig.~\ref{fig:180}(a-b) that both $P_t$ and
$P_r$ components are strongly non-zero, and the rotation of the
polarization as one progresses through the domain wall, which
is the characteristic feature of a Bloch-type wall, is clearly
visible.  It is also obvious that this Bloch FDW resembles an
adjacent pair of 71$^\circ$ and 109$^\circ$ FDWs.  We also find
that this Bloch wall has a significantly lower energy than that
of the Ising-like wall of Fig.~\ref{fig:180ising}, supporting the
conclusion that the Bloch solution is the global minimum for the
case of the R180$^\circ$\{1$\bar{1}$0\} FDW.

We have also carried out corresponding simulations of the
Ising-like and Bloch configurations of the R180$^\circ$\{1$\bar{1}$0\}
FDW using the GLD model.  If this is done using the full strength
of the gradient term under periodic boundary conditions, we find
a stable Bloch solution only when the centers of the domain walls can be
at least 3.4\,nm apart.  This is equivalent to using a supercell
of 120 atoms in the first-principles calculations, and imposing
the rhombohedral epitaxial strain does not change this result
very much.  On the other hand, we find that if the gradient term
is reduced by about 40\%, then stable solutions become possible
under the same periodic boundary condition as in the 80-atom supercell.
The polarization profiles calculated for the R180$^\circ$\{1$\bar{1}$0\}
FDWs using the GLD model with reduced gradient term,
with the same domain-wall distance and elastic boundary
conditions as in the first-priciples calculations,
are shown in
Figs.~\ref{fig:180ising}(c-d) and \ref{fig:180}(c-d) for the
Ising-like and Bloch cases respectively.
The GLD results are clearly now in good qualitative agreement with
the first-principles calculations.

\begin{table}
\caption{\label{tab:walls}%
(Color online)
Summary of computed energies and widths of R71$^\circ$, R109$^\circ$, and
R180$^\circ$\{1$\bar{1}$0\} FDWs in BaTiO$_3$.
LDA indicates first-principles results; GLD and GLD-r refer to the
Ginzburg-Landau-Devonshire model with original and reduced gradient
term, respectively.}
\begin{ruledtabular}
\begin{tabular}{lcccccccc}
&& \multicolumn{3}{c}{Wall width (nm)} &&
   \multicolumn{3}{c}{Wall energy (mJ/m$^2$)} \\
&& \;LDA\; & \;GLD\; & \;GLD-r && \;LDA\; & \;GLD\; & GLD-r \\
\hline
Ising R71$^{\circ}$  && 0.33 & 0.58 & 0.37 && 3.8 & 5.0 & 3.2\\
Ising R109$^{\circ}$ &&0.36 &0.54&0.34 &&11.1&10.6&6.8\\
Bloch R109$^{\circ}$ &&1.01&1.10&0.72  &&11.2&10.2&6.2\\
Ising R180$^\circ$   && 0.51 & -- &0.83     &&26.6& -- &27.2\\
Bloch R180$^\circ$  &&1.38 & -- &1.40   &&24.0 & -- &27.6\\
\end{tabular}
\end{ruledtabular}
\end{table}

The computed energies and widths of all of the FDWs are collected
and presented in Table \ref{tab:walls}.  It is evident that the
first-principles and GLD results are in broad agreement.  As expected,
the GLD model with reduced gradient term yields narrower walls and
lower wall energies, yielding improved agreement for the Ising
R71$^\circ$ and R109$^\circ$ cases, but somewhat overshooting for
the Bloch R109$^\circ$ case.

For the R180$^\circ$ FDWs, the values given in Table~\ref{tab:walls}
ought not be taken too seriously because the repeat
distance of the supercell is rather short compared to the FDW width.
In fact, we did not succeed in finding stable FDW solutions with
the original GLD model.  The present 80-atom supercell is large enough to
give stable solutions in both the LDA and reduced-GLD calculations,
but their properties are undoubtedly not yet converged with supercell
size.  The FDW energy can be expected to fall with supercell size,
so the energies in Table~\ref{tab:walls} should be taken as
upper bounds.  The ``crowding'' of the Bloch R180$^\circ$ walls
appears to be more serious than for the Ising ones, which
can explain why the reduced GLD model predicts a slightly higher energy
for the Bloch compared to the Ising FDW.  When the GLD calculations
are repeated for larger FDW separations, they clearly predict
that the Bloch configuration is lower in energy.\cite{Marton}
The first-principles calculations already predict the Bloch wall
to be lower in energy for the 80-atom supercell, and this trend
would only be strengthened if we could afford to repeat the
calculations at increasingly larger FDW separations such that
well-developed rhombohedral domains could form between domains.

\section{Discussion}
\label{sec:discuss}

\begin{figure}
\caption{\label{fig:sym}
(color online)
The path swept by the tip of the polarization vector
while rotating from ${\bf P}_1$ to ${\bf P}_3$ when crossing through
hypothetical Bloch-type FDWs for the R71$^\circ$, R109$^\circ$, and
R180$^\circ$\{1$\bar{1}$0\} geometries.  For the R71$^\circ$
case the path (which rotates in a plane parallel to the FDW)
is shown in red to indicate that it is unfavorable.}
\includegraphics[width=3.4in]{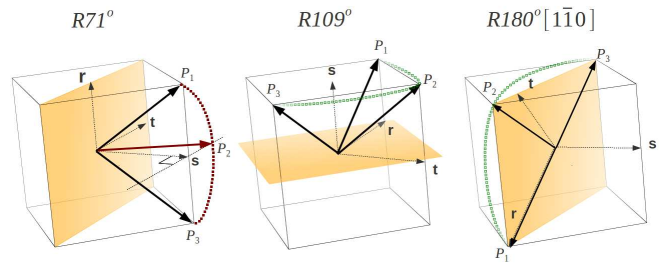}
\end{figure}

Among the three investigated FDWs, the polarization vector rotates by
the smallest angle in the R71$^\circ$ FDW, and by the biggest angle in the
R180$^\circ$\{1$\bar{1}$0\} FDW, so it is not surprising that the former
has the smallest energy and the latter the biggest, as summarized in
Table~\ref{tab:walls}.
We can also propose a simple explanation for the fact that the
first-principles calculations and the GLD model results predict
an Ising nature for the R71$^\circ$ FDW, a Bloch nature for the
R180$^\circ$\{1$\bar{1}$0\} FDW, and a very small energy difference
between Ising and Bloch solutions for the R109$^\circ$ FDW. The
path the polarization vector would take in rotating from one
side of the wall to the other in a Bloch-type solution is shown
for each of these FDWs in Fig.~\ref{fig:sym}.
In a hypothetical Bloch R71$^\circ$ FDW the polarization vector would pass
close to the center of one of the adjacent faces, which corresponds to
a tetragonal polarization state and is not energetically favorable
in the rhombohedral phase.

On the other hand, the Bloch 109$^\circ$ FDW can be considered
as a combination of two R71$^\circ$ FDWs. As can be seen from
Table~\ref{tab:walls}, the total energy of two Ising R71$^\circ$
FDWs is rather close to the energy of one Ising 109$^\circ$ FDW,
from both the LDA and GLD calculations.
By way of a caveat, we point out that such a comparison
may be overly simplistic because the Ising walls comprising
the Bloch 109$^\circ$ FDW experience a foreign strain environment
and do not conform to the plane of mechanical compatibility of
a true R71$^\circ$ FDW.  Nevertheless, the comparison does hint
that we should not be surprised to find
the Ising and Bloch configurations to be competitive here.
Finally, for the Bloch 180$^\circ$\{1$\bar{1}$0\} FDW, which can
be regarded as one R71$^\circ$ FDW plus one R109$^\circ$ FDW, the
LDA calculations indicate that the sum of the energies of these
two walls is much lower than that of a single Ising 180$^\circ$
FDW.
The above caveat has perhaps even more force here, but again
we can roughly understand in these terms
why the R180$^\circ$\{1$\bar{1}$0\} FDW can only adopt a Bloch form.

\section{Summary}
\label{sec:summary}

In conclusion, we have calculated the domain wall widths, energies,
and polarization profiles for Ising and Bloch ferroelectric domain
walls in the zero-temperature rhombohedral phase of BaTiO$_3$ using
both first-principles and Ginzburg-Landau-Devonshire methods.
The first-principles results confirm the expectation that
180$^\circ$ domain walls are of Bloch type, adopting a configuration
resembling a pair of 109$^\circ$ and 71$^\circ$ walls in close
proximity.  For the case of the  109$^\circ$ wall, Ising and
Bloch configurations are competitive.  The Ginzburg-Landau-Devonshire
results are brought into improved agreement with the first-principles
calculations if the coefficient of the gradient term is
reduced by about 40\%.  In view of the uncertainties in the
original extraction procedure for the coefficients, it is not
surprising that these parameters can be improved; indeed, it is
encouraging that even the original parameters gave qualitatively
sound results.

While we have not extended our work to other rhombohedral
ferroelectrics such as KNbO$_3$, we expect that these may show
a similar pattern of behavior.  When instabilities other than
ferroelectric ones are also present, the domain-wall behavior
can become more complicated, as for example with the
octahedral rotations and magnetic ordering that play a role in
BiFeO$_3$.\cite{Lubk} However, we hope that the present work
will serve as a useful benchmark for domain walls in rhombohedral
ferroelectrics generally, and will lead to an improved understanding
of ferroelectric domain dynamics and switching in these systems.

\section*{Acknowledgments}

This work was supported by ONR Grant N00014-05-1-0054 and by
Project GACR P204/10/0616 of the Czech Science Foundation.


\end{document}